*Original research*

# 3D printable strain rate-dependent machine-matter

S. Janbaz[a,b,1,*], D. Fan[c,*], M. Ganjian[a], T. van Manen[a], U Staufer[c], A.A. Zadpoor[a]

[a] *Department of Biomechanical Engineering, Faculty of Mechanical, Maritime, and Materials Engineering, Delft University of Technology, Mekelweg 2, 2628CD, Delft, The Netherlands*
[b] *Institute of Physics, Universiteit van Amsterdam, Science Park 904, 1098 XH Amsterdam, The Netherlands*
[c] *Department of Precision and Microsystems Engineering, Faculty of Mechanical, Maritime, and Materials Engineering, Delft University of Technology, Mekelweg 2, 2628CD, Delft, The Netherlands*
[1] Corresponding author, s.janbaz@uva.nl, *Equal contributions

Machine-matter, of which mechanical metamaterials and meta-devices are important sub-categories, is emerging as a major paradigm for the design of advanced functional materials. Various exciting applications of these concepts have been recently demonstrated, ranging from exotic mechanical properties (*e.g.*, negative and extremal properties) to device-like (*e.g.*, mechanical logic gates) and adaptive (*e.g.*, shape-shifting) functionalities. The vast majority of the studies published to date have, however, focused on the quasi-static behavior of such devices, neglecting their rich dynamic behavior that could be harnessed to bring about far greater levels of sophistication. Recently, we proposed a new class of strain rate-dependent mechanical metamaterials that are made from bi-beams (*i.e.*, two identical laterally-attached beams with vastly different degrees of strain rate-dependency). The buckling direction of such bi-beams can be reliably controlled with the applied strain rate. The proposed approach, however, suffers from a major limitation: 3D printing of such bi-beams with such a 'strong' differential strain rate-dependent response is very challenging, if not impossible. Here, we propose an alternative approach that only requires a 'weak' differential response and a rationally designed geometric artifact to control the buckling direction of bi-beams. We present an analytical model that describes the landscape of all possible combinations of geometric designs and hyperelastic as well as viscoelastic properties that lead to the desired strain rate-dependent switching of the buckling direction. We support our analytical model with an extensive computational analysis of the post-buckling behavior of the proposed designs. We also demonstrate how two 3D printing techniques, namely multi-material polymer jetting and single-material two-photon polymerization, can be used to fabricate the proposed bi-beams with microscale and submicron resolutions, respectively. More importantly, we demonstrate how the requirement for a weak differential response eliminates the need for multi-material 3D printing, as the change in the laser processing parameters is sufficient to achieve a high enough level of differential response. Finally, we use the same 3D printing techniques to produce strain rate-dependent gripper mechanisms as showcases of potential applications.

## 1. INTRODUCTION

The boundary between materials and devices is becoming blurred, as advanced functional materials are increasingly exhibiting properties that are usually ascribed to devices. Within the context of mechanical design, this trend has led to the emergence of "machine-matter", where architected materials exhibit some or most of properties and functionalities that have been traditionally imputed to machines. On the one side of the spectrum, mechanical metamaterials[1-3] are rationally designed to exhibit exotic material-like properties, such as negative effective values of the Poisson's ratio[4-6], thermal expansion[7-9], and stiffness[10,11]. On the other side of the spectrum, however, one finds such concepts as mechanical logic gates[12,13], adaptive-stiffness mechanisms[14,15], and shape-shifting designs[16-19], which exhibit device-like functionalities.
As opposed to traditional machines, which are usually highly dynamic, the vast majority of the mechanical metamaterials designed to date work quasi-statically. The quasi-static nature of such designs is a major limiting factor that prevents mechanical metamaterials from benefiting from the rich physics of dynamic designs particularly those based on viscoelasticity. A few recent studies[20-23] have started to tap into the huge potential of viscoelastic mechanical metamaterials. One such study[20] proposed the concept of bi-beams as structural elements that could be used to create strain rate-dependent mechanical metamaterials. Such mechanical metamaterials could achieve unique properties, including negative viscoelasticity and strain rate-dependent switching between an auxetic and a conventional response. Bi-beams are made by laterally attaching two geometrically identical beams whose strain rate response is highly different. While one of the beams (*i.e.*, the hyperelastic beam) is largely strain rate insensitive, the other beam (*i.e.*, the visco-hyperelastic beam) is highly sensitive to the applied strain rate. Combining two such beams in a bi-beam construct allows for a reliable strain rate-dependent control of the buckling direction. This switchable buckling direction is the principle underlying all the unique properties of such mechanical metamaterials.

A major limitation of this bi-beam concept is the need for materials with 'strong' differential strain rate-dependent responses. It is very challenging, if not impossible, to 3D print two materials with such a high degree of differential strain rate-dependent response. Indeed, most polymers show significant levels of viscoelasticity, making it difficult to find suitable material candidates for the fabrication of the hyperelastic beam. Given the fact that 3D printing is the only practical approach for the fabrication of complex architected materials, any design that cannot be 3D printed, while theoretically interesting, is intrinsically limited in terms of its potential impact.

Here, we propose an alternative bi-beam concept that only requires a 'weak' differential strain rate-dependent response to control its buckling direction, making it easily 3D printable. In the case of a bi-beam made by attaching a 'hyperelastic' beam to a 'visco-hyperelastic' beam (*i.e.*, our previous design), the strain rate-dependent switching behavior was caused by a competition between the local buckling modes of the hyperelastic and visco-hyperelastic beams that effectively behaved as two beams resting on elastic foundations[20,24]. The bi-beams proposed here are, however, made of two 'visco-hyperelastic beams (*i.e.*, beams 1 and 2) (Figure 1A). Due to the relatively weak differential strain rate-dependent response, the mechanistic routes exploited in the previous design break down, meaning that material properties alone cannot achieve the desired strain rate-dependent switching of the buckling direction. We, therefore, introduced a rationally designed and purposefully applied geometric artifact in the form of a curvature with a radius, $R$, to the shape of the beams (Figure 1). The combination of this geometric artifact and the weak differential response makes it possible to achieve strain rate-dependent control of the buckling direction of the bi-beams that make up our machine matter. An important consequence of the proposed concept is that there may be no need for a multi-material 3D printing technique, as the differential mechanical properties created when changing the printing parameters of a single material are often sufficient for creating the desired strain rate-dependent behavior. We present an analytical and a computational model to explain the proposed theoretical concept and to rationally design the required geometric artefacts. We then realize the proposed (micro) bi-beams using both a multi-material (*i.e.*, polymer jetting) and a single-material (*i.e.*, two-photo polymerization) 3D printing techniques. Finally, we use the micro bi-beams for the design and fabrication of a piece of machine matter in the form of a gripping mechanism.

## 2. RESULTS AND DISCUSSION

Let us consider a bi-beam made of two different visco-hyperelastic materials where the shape of the bi-beam is an arc of a large circle (*i.e.*, $R/2L \gg 1$) (Figure 1A). Strain rate-dependent metamaterials[20] where a 'strong' differential strain rate-dependent response is required is a special case of this design. In that special case, the radius of the circle is infinitely large (*i.e.*, both beams are fully straight). Moreover, the beam whose response is (almost) strain rate independent (*i.e.*, hyperelastic beam), has a higher elastic modulus than the other beam (*i.e.*, visco-hyperelastic beam). Such a bi-beam construct always buckles in the direction of the stiffer constituting beam[20] (Figure 1B). For small strain rates, that stiffer beam is the hyperelastic beam. For high enough strain rates, the apparent elastic modulus of the visco-hyperelastic beam exceeds that of the hyperelastic beam, causing the bi-beam to buckle in the direction of the visco-hyperelastic beam (Figure 1B).

Bi-beams with 'weak' differential strain rate-dependent response work differently (Figure 1C) because the difference in the strain rate dependency of both materials may not be high enough to change the order of the stiffness of the beams. A rationally designed and purposefully introduced geometric artifact (*i.e.*, a finite $R$) is, therefore, required. The geometric artifact creates the tendency for the bi-beam to buckle to a specific direction (left in the case of the geometry show in Figure 1A). As we will see later (in Figure 1F), there are multiple combinations of hyperelastic and viscoelastic properties that could lead to the strain rate-dependent switching behavior of whichever one has been selected for a more detailed analysis in Figure 1C. In this scenario, the material on the right is so much stiffer than the material on the left that the bi-beam overcomes that geometric tendency and buckles to the right for the small values of the strain rate. For very high values of the strain rate, on the other hand, the stiffness of the material on the left increases more than that on the right. As a consequence, the higher stiffness of the material on the right is not high enough to overcome the effect of the geometric artifact. The bi-beam, therefore, buckles to the left (Figure 1C).

To gain insight into the most energetically favorable buckling modes of bi-beams with 'weak' differential strain rate-dependent responses, we developed an analytical model (see Materials and Methods). The finite deformations of the materials were captured using two Neo-Hookean constitutive equations with the long-term materials constants $C_{10,1}^\infty$ and $C_{10,2}^\infty$ corresponding to beams 1 and 2, respectively. The geometric artifact favors buckling towards the beam made from material 1. The viscoelastic behaviors of the beams were described using two one-term Prony series with the dimensionless Prony coefficients $g_{1,1}$ and $g_{1,2}$ and the relaxation times $\tau_{1,1}$ and $\tau_{1,2}$ where the latter subscripts 1 and 2 refer to beams 1 and 2, respectively. The analytical analysis was focused on finding the geometrical designs and material combinations for which the strain rate-dependent switching of the buckling direction was possible, provided that the strain rate differential was high enough. To determine those limits, we considered the two boundary-defining cases of extremely high (*i.e.*, $\dot{\varepsilon} \to \infty$) and extremely low (*i.e.*, $\dot{\varepsilon} \to 0$) strain rates. Under such conditions, the analytical model predicts that there are only two material-related parameters $\alpha = C_{10,1}^\infty/C_{10,2}^\infty$ and $\beta = (1 - g_{1,2})/(1 - g_{1,1})$ and one geometry-related parameter $c/2a$ ($c = R - \sqrt{R^2 - L^2}$), corresponding to the equivalent curved configuration of bi-beams illustrated in Figure 1D, that determine whether or not the strain rate-dependent switching of the buckling direction takes place for a bi-beam with given values of the length ($2L$) and width ($a$). The master curve relating the geometric and material designs to each other is presented in Figure 1E. From this curve, it is clear that there is a cut-off value of $c/2a = 0.3$ for the geometric designs that could be used for the strain rate-dependent control of the buckling direction. This cut-off value is valid regardless of the material properties of the beams constituting the bi-beam and indicates that the curvature of the geometric artifact may not go beyond a certain threshold if the differential material response is to overcome the geometric tendency to buckle towards a specific direction (*i.e.*, right at low strain rates in Figure 1C). The predictions of the analytical model agree with our computational results (Figure 1F). For example, for a $c/2a$ value of 0.53, no strain rate-dependent switching behavior is predicted by our computational models (Figure 1C) while such a behavior is seen for all the $c/2a$ values below 0.3 (Figure 1F, insets *i* to *iv*). For values between 0 and 0.3, there are specific values of the material parameter $\alpha\beta$ for which the strain rate-dependent switching of the buckling direction is possible, provided that the applied strain rate is high enough. This parameter describes the relationship between the elastic and viscous properties of the beams for which the switching is possible. In general, the more nuanced the geometric artifact (*i.e.*, smaller $c/2a$), the higher the required value of the material parameter $\alpha\beta$. Furthermore, the parameter $\alpha$ determines the buckling direction at low strain rates while the parameter $\alpha\beta$ does the same at high strain rates. For example, for a negligible beam curvature (*i.e.*, $c/2a \approx 0$), switchability requires that $\text{sgn}(\alpha - 1) \neq \text{sgn}(\alpha\beta - 1)$, meaning that if the ratio of the stiffness values at low strain rates is below 1, the one corresponding to high-strain rates should exceed 1, and vice versa. Regarding the elastic *vs.* viscous properties, a double-triangle region (in the logarithmic-logarithmic plane) defines the combinations of the elastic and viscoelastic properties for which the switching is possible. This double-triangle consists of two regions: the yellow region in which the bi-beam buckles to the geometrically favored direction (*i.e.*, left in Figure 1F) for very low values of the strain rate and the blue region in which the switching is in the opposite directions (*i.e.*, right in Figure 1F). The working mechanism within the blue region is the same as described for weak viscoelasticity in the previous paragraph (Figure 1C). While in the blue region, $\alpha$ is always below unity, it can be either above or below unity in the yellow region, meaning that the beam on the right may have a higher or a smaller value of stiffness for very small values of the strain rate. Within the yellow region, the bi-beam always and regardless of the $\alpha$ value buckles to the left for very small values of the applied strain rate. For very high values of the strain rate, however, the stiffness of the beam on the right increases so much more than that on the left that it can overcome the effects of the geometric artifact and the bi-beam buckles to the right. The $\beta$ value should, therefore, be below 1 (*i.e.*, $g_{1,2} > g_{1,1}$), which would mean that the stiffness of the beam on the right increases more than that on the left, as the applied strain rate increases. When $\alpha > 1$, the beam on the left is initially (*i.e.*, for very small values of the strain rate) stiffer than on the right. Given that the buckling to left is geometrically favored as well, the bi-beam buckles to the left for very small values of the applied strain. When the strain rate is very high, the stiffness of the beam on the right should increase so much that it can overcome both the geometric tendency to buckle to the left and the stiffness deficit with the left side. That is why the $\beta$ value should be very low for very high values of $\alpha$ (*e.g.*, in insets *ii* and *iii* of Figure 1F). As the curvature of the geometric artifact increases, the double-triangle region shifts towards lower $\alpha$ values. For high enough curvatures of the geometric artifact, the strain rate-dependent switching is only possible for $\alpha < 1$ (*e.g.*, inset *iv* in Figure 1F). That is because the geometric effect is so strong that the additional stiffness generated through viscoelastic effects cannot overcome the geometric tendency to buckle to the left. In terms of strong *vs.* weak differential viscoelastic response, a weak differential response is sufficient for the blue region as well as for the yellow region with $\alpha < 1$ while a strong differential response may be needed for the yellow region with $\alpha > 1$. In all cases, the predictions of our linearized analytical model agree with those obtained using our full visco-hyperelastic finite element models that took the full geometry of the bi-beams into account (Figure 1F), indicating that the linearized analytical model is successful in capturing the most important features of the strain rate-dependent buckling behavior of the designed bi-beams (Figure 1F, insets *i* to *iv*).

To demonstrate the possibility of the proposed designs, we used multi-material polymer jetting and two-photon polymerization (2PP) to fabricate bi-beams at the macro- and microscales (Figure 2, see Materials and Methods). In the case of the macroscale specimens, the 'weak' differential strain rate-dependent response was achieved using two types of polymers (*i.e.*, Tango Plus and Agilus) with slightly different mechanical behaviors (Figure 2A). In the case of the microscale specimens, we used different levels of the polymerization laser power to create the required differential response between the beams. Two types of materials, namely PDMS (polydimethylsiloxane) and IP-Dip (a commercially available acrylate-based resin), were used, both of which exhibited the expected 'weak' differential strain rate-dependent response (Figure 2A, C). In all cases, the fabricated specimens buckled in the desired direction determined by the applied strain rate (Figure 2B, D). These findings demonstrate the high potential of the proposed design approach, given that the applied 3D printing techniques are some of the most versatile and powerful techniques currently available. In the case of multi-material polymer jetting, a wide range of materials are commercially available that can be used to create architected materials (*e.g.*, lattice structures) using the 3D printable bi-beams demonstrated here. As for 2PP, there are two important points that make it particularly useful for fabricating bi-beams. First, 2PP makes it possible to create the required weak differential strain rate-dependent response simply by changing the printing parameters and without a need for a second material. Second, 2PP offers submicron printing resolutions (*i.e.*, 200 nm), which could be used to create architected materials at the nano-/microscales.

Focusing on the manufacturability and reliability of microscale bi-beams, we studied the effects of the geometry (*i.e.*, radius of curvature), printing material, and printing parameters on the strain rate-dependent buckling behavior of microscale bi-beams. Three series of specimens with gradually decreasing dimensions (*i.e.*, full-size, half-size, and quarter-size) were manufactured from two different materials (*i.e.*, PDMS and IP-Dip) while maintaining the same aspect ratio (6.4), print slicing (300 nm), and hatching size (200 nm) (Figure 3). When the microscale bi-beams were made from PDMS, switching in the buckling direction was observed only for the full size PDMS bi-beams ($L = 160$ μm) for a radius of curvatures $R = 2$ and 3 mm. The details of the microscale bi-beams could not, however, be very accurately reproduced and a significant number of specimens contained defects (Figure 3A). The strain rate-dependent buckling behavior, while present, was not very predictable and no clear boundary was found between 'low strain rate' and 'high strain rate' behaviors. This is most likely due to the limitations of the 2PP technique in producing defect-free, geometrically precise microscale bi-beams from the PDMS resin used here.

In comparison to the PDMS resin, IP-Dip is a well-optimized photoresist that can be used to fabricate high-resolution microscale bi-beams. We, therefore, chose three different dimensions (*i.e.*, $L = 160, 80, 40$ μm) corresponding to the full-, half-, and quarter-sized specimens to evaluate their manufacturability and the effects of the printing parameters on the strain rate-dependent buckling behavior of the resulting microscale bi-beams (Figure 3B, slicing distance = 300 nm, hatching distance = 200 nm in all cases). The full- and half-size bi-beams exhibited predictable and reproducible strain rate-dependent switching in their buckling direction for specific radii of curvature (*i.e.*, $R = 12$ mm for the full-size and $R = 2.5$ mm for the half-size microscale bi-beams). The unequal ratios of the height to the radius of curvature that are required for the robust switching of the buckling direction in the full-size and half-size bi-beams highlights the importance of the printing resolution (*i.e.*, slicing and hatching) on the regulation of the buckling behavior of microscale bi-beams. For smaller dimensions (*i.e.*, quarter-size bi-beams), the switching behavior, while present, was not predictable (Figure 3B). That is most probably a consequence of the limited printing resolution that distorts the geometry of the specimens and prevents them from scaling with full detail. The most important geometrical detail that is lost as smaller dimensions are approached are the cut-outs incorporated into the connection of the bi-beams to the clamping sites at the top and bottom of the microscale bi-beams. Given the importance of the cut-outs in defining the boundary conditions of the bi-beams, their imprecise fabrication leads to unpredictable buckling behavior.

To better understand the size effects, we measured the strain rate-dependent properties of the bulk materials photopolymerized using 40% and 45% laser power. In order to evaluate the effects of the geometry and printing resolution on the strain rate-dependent properties of IP-Dip, we used three different types of geometries and scaled them to create prisms with three different heights (*i.e.*, 60, 40, and 20 μm). The compression tests of the bulk materials confirmed the meaningful strain rate-dependency of the IP-Dip polymers manufactured using 40% and 45% laser powers (Figure 4A). It is clear that a higher laser power results in stiffer materials. Looking at the stiffness of the bulk materials fabricated using the high and low values of the laser power, the ratio of the stiffness of high-power materials to that of the low-power ones generally decreases for the higher values of the strain rate. For example, the ratio of the stiffness of the specimens with a square cross-section (height = 60 μm) is 24% lower for the higher strain rates as compared to the stiffness ratio at a lower strain rate. It is important to mention that the stiffness values were measured between 30% and 90% of the maximum strain values of the corresponding stress strain curves. The drop in the stiffness ratio for the higher strain rates demonstrates the effectiveness of our assumptions for programming the strain rate buckling behavior of the microscale bi-beams made from IP-Dip. It is, however, clear that the stiffness values are size-dependent and that the geometry may also influence the rate-dependent mechanical properties of IP-Dip materials to some extent (Figure 4A).

Furthermore, we experimentally analyzed the effects of the printing resolution on the buckling behavior of the quarter-size IP-Dip bi-beams. We first fixed the slicing size (layer thickness) to 300 nm (similar to the structures presented in Figure 2) and evaluated the effects of the hatching size (the offset in the laser line tracking) and the laser power on the strain rate-dependent buckling behavior of the specimens with a radius of curvature equal to 3 mm (Figure 4B). The experiments revealed the dependence of the buckling behavior of the microscale bi-beams both on the hatching size and the photopolymerization power: low-power photopolymerization is more effective in amplifying the differential strain rate-dependent response when the hatching size is small while larger hatching sizes are required for high-power photopolymerization. This is consistent with the expected effects of a higher laser power in over-curing the previously printed IP-Dip, which is diminished when a larger hatching size is used. Furthermore, we evaluated the influence of the printing resolution on the scalability of the geometrical details of full- and quarter-size microscale bi-beams. High resolution SEM images of the full- and quarter-size bi-beams clearly show that the resolution of the printing process plays an important role in the precise reproduction of the geometrical details, such as the cut-outs present at the clamping sites, that, if not precisely realized, could diminish the robustness of the strain rate-dependent buckling behavior of the microscale bi-beams (Figure 4C). Such an imprecise geometry is clearly visible in the case of the quarter-size bi-beams once the resolution of printing (the hatching = 200 nm and slicing = 300 nm) is in the range of the dimensions of the cut-outs present at the either end of the bi-beams.

To eliminate the geometrical imperfections from our evaluations, we fabricated quarter-size bi-beams with an increased distance between the pedestal and the top side of the cut-out gaps. This ensured that the desired shape and, thus, boundary conditions of the bi-beams is maintained as they are scaled down in size. The new quarter-size microscale bi-beams were manufactured using four different combinations of the laser power. The radius of curvature in all the quarter-size bi-beams was 3 mm. For the lower values of the laser power (*i.e.*, 35 and 40), the strain rate-dependent buckling behavior was only minimally predictable. Increasing the laser power for both beams constituting the microscale bi-beams reveals that a simultaneous rise in the hatching and slicing distances transforms the left buckling into right buckling while switching behavior is observed at the interface of the left and right buckling zones (Figure 4D). Further increase in the laser powers confirms the impractical programming of strain rate-dependency of the buckling behavior of quarter-size micro bi-beams. That may be because a high laser power strongly solidifies the IP-Dip resin, lowering its viscoelasticity. More difference in the laser power (35/45) confirms the observations for (40/45) laser power set while due to the lower long-term stiffness (or strong differential viscoelastic properties of the layers) of the left layer (laser power 35) the switching was observed for the highest resolution (Figure 4E).

If the level of compaction and the rate of the applied compressive strains exceed certain thresholds, the microscale bi-beams will be damaged. Inspecting the full-size bi-beams revealed that a crack was initiated from right to left at the clamping sites of the bi-beams showing left buckling, compressed at high strain rates (Figure 4E). This is expected given the local stress concentrations, high strain rates, and large deformations. The direction of the crack was found to be from left to right when right buckling happened at low speeds. Reducing the purposefully introduced curvature of the bi-beams can prevent the initiation of the cracks when compressing them at low speeds (Figure 4E), most likely by reducing the stress concentration. As such permanent damage and possible plastic deformation prevent the recovery of the bi-beams after the first compaction cycle, the direction of buckling becomes permanently independent from the rate of the applied strain in the successive loading cycles. That means the permanent direction of buckling can be programmed based on the first loading pattern. Interestingly, we observed that this memory can be erased by simply annealing the undamaged bi-beams compressed at a low strain rate at a temperature of 70 °C for one hour. The existence of a crack renders the annealing ineffective in the case of the damaged bi-beams compressed at high strain rates.

Finally, we designed compliant grippers to showcase the potential of the proposed designs for the development of machine-matter. In this design, both arms of the compliant grippers are carried using symmetrically aligned parallel links, forming four-bar mechanisms with flexible joints at the either side of the arms (Figure 5A). Bi-beams were used to push or pull the arms of the gripper to the left or the right, resulting in an open or a close configuration. To demonstrate the manufacturability and scalability of the illustrated concept, we used the same additive manufacturing techniques as before (*i.e.*, Polyjet and 2PP) to create both macroscale bi-beams made from TangoPlus and Agilus (Figure 5B) and microscale bi-beams made from IP-Dip (Figure 5B). In the case of the macroscale specimens, a voxel-based combination of Agilus and Vero Black was used to improve the stiffness of the materials and prevent the compliant joints from buckling. The macroscale bi-beams behave similarly to the specimen presented in Figure 2B, buckling to the left at low speeds (*e.g.*, 20 mm/min) and opening the gripper. At high speeds (*e.g.*, 950 mm/min), the bi-beams buckle to the right, causing the grasping mechanism to close (Figure 5B). As for the microscale specimens, the right buckling of full-size bi-beams closes the gripper at low speeds (*e.g.*, 0.1 µm/s) and opens it at high speeds (*e.g.*, 250 μm/s). The high level of scalability and the feasibility of manufacturing complex, hybrid mechanisms are promising for the development of miniaturized soft robotic devices, such as steerable surgical tools that exhibit multiple modes of intricate deformations and can be controlled by one single wire, where the speed determines the pattern of actuation. Another example would be protective exo-suits that are comfortable to wear but protect the body from low-energy fractures (*e.g.*, due to osteoporosis) when subjected to high strain rate events, such as the impact caused by stumbling and falling. Exposure to high strain rate events also increases the stiffness of the beams due to the viscoelastic nature of the underlying materials, thereby increasing their load bearing capability (Figure 5C).

### 3. OUTLOOK AND CONCLUSIONS
In summary, we proposed a design paradigm based on a 'weak' differential strain rate-dependent response of the bi-beams to enable 3D printing of strain rate-dependent machine matter using the existing additive manufacturing technologies both at the macro- and microscale. The presented analytical model and computational results clearly show the underlying mechanisms and how various geometrical and material parameters determine the strain rate-dependent buckling direction of the bi-beam. The role of purposefully introduced geometric imperfections is particularly interesting, as it is the main mechanism that allows us to relax the requirement regarding a 'strong' differential strain rate-dependent response in the two beams constituting a bi-beam and replacing it by a mere 'weak' differential response requirement. Such a 'weak' differential response is much more likely to be achieved using the existing 3D printing techniques particularly given the fact that most of the existing technologies are single-material. As a consequence of this relaxed requirement, we could manufacture microscale bi-beams using 2PP, simply by changing the laser power and without a need for a second material. Moreover, the proposed approach does not require the stretchability and flexibility of previously proposed elastomeric bi-beams, meaning that the bi-beams can be made of non-stretchable polymers. Advanced device-like materials can, therefore, be designed using this approach for different types of applications ranging from micro-machines and mechanical calculators to surgical meta-devices. As an example, we showed that one bi-beam is sufficient to effectively alter the actuation of a (micro-) gripper in response to the applied strain rate. The micro-gripper made using 2PP is small enough to be able to capture a single human cell and release it at a determined site. Likewise, multiple rationally designed bi-beams can be organized in a device-like material to perform much more complex strain rate-dependent functions, such as those required for the development of logic gates. The downside of a 'weak' differential strain rate-dependent response is that the observed switching in a buckling direction is less robust as compared to the extremely robust response observed in the bi-beams made with a 'strong' differential response[20]. However, as our repertoire of 3D printable high-performance, low-viscoelasticity polymers expands, the robustness of the bi-beams made using the 'weak' differential response is expected to approach that of the bi-beams made using the 'strong' differential response approach.

### 4. MATERIALS AND METHODS

#### 4.1. Analytical model
We built an analytical model to evaluate the effects of geometry (*i.e.*, dimensions and the radius of curvature) and material properties on the buckling direction of bi-beams made of two visco-hyperelastic beams. Assuming that the cut-outs incorporated around the clamping sites of the bi-beams effectively behave as pivot connections, we formulated the bending of the bi-beams using the Euler-Bernoulli beam theory as:

$$EI \frac{d^2w}{dx^2} = M(x) \tag{1}$$

A curvature is introduced into the geometry of the beams. The bending moment caused by the compressive force is, therefore, a function of the geometry of the beams and varies along their length (Figure 1A). The bending moment is given as $M(x) = P(\Delta - y_0(x))$, where $\Delta$ is the shift of the neutral axis from the interface to the stiff side and $y_0(x)$ is the offset of the axial load with respect to the interface of both beams at a distance $x$ from the middle of the bi-beams. Considering that the geometries of the beams are curved, the perpendicular distance between the axial load and a point on the 'neutral line' can be approximated as $y_0 = c - \frac{c}{L^2}x^2$, where $c = R - \sqrt{R^2 - L^2} \ll L$, $R$ is the radius of curvature of the purposefully introduced geometrical imperfection, and $L$ is the net half-length of the bi-beams (*i.e.*, excluding the cut-out regions). Substituting this relationship in Equation (1) yields:

$$\frac{d^2w}{dx^2} = \frac{P}{EI}\left(\Delta - c + \frac{c}{L^2}x^2\right) \tag{2}$$

We can, then, calculate the lateral deflection of the bi-beam as:

$$w(x) = \frac{P}{EI}\int_0^x \int_0^x \left(\Delta - c + \frac{c}{L^2}x^2\right) dx^2 = \frac{P}{EI}\left(\frac{1}{2}\Delta x^2 - \frac{1}{2}cx^2 + \frac{c}{12L^2}x^4 + C_1 x + C_2\right) \tag{3}$$

By applying the boundary conditions $\frac{dw}{dx}(0) = 0 \; and \; w(0) = 0$, the constants $C_1 = 0$ and $C_2 = 0$ can be determined, and we can write:

$$w(L) = \frac{PL^2}{EI}\left(\frac{1}{2}\Delta - \frac{5}{12}c\right) \tag{4}$$

The critical switching behavior of the bi-beam, therefore, corresponds to the state that the movement of the pivot in the lateral direction is zero:

$$\Delta_{cr} = \frac{5}{6}c \tag{5}$$

The offset of the neutral axis from the interface of the beams with equal thickness ($a/2$) can be calculated as a function of the elastic moduli of both constituting beams of the bi-beam (Figure 1A):

$$\Delta = \frac{\frac{1}{4}aE_2\frac{a}{2} - \frac{1}{4}aE_1\frac{a}{2}}{\frac{1}{2}a(E_1 + E_2)} \tag{6}$$

By defining $A = \frac{E_1}{E_2}$, $\Delta$ can be re-written as:

$$\Delta = \frac{a}{4}\frac{1-A}{1+A} \tag{7}$$

Substituting Equation (7) in Equation (5), the critical ratio of the elastic moduli is determined as:

$$A^* = \frac{\frac{a}{4} - \frac{5c}{6}}{\frac{5c}{6} + \frac{a}{4}} \tag{8}$$

If $A < A^*$, the value of $w(L)$ is positive and the bi-beam buckles to the right (*i.e.*, the negative direction of the $y$ axis in Figure 1A). For $A > A^*$, the bi-beam buckles to the left.

For visco-hyperelastic beams, where the viscous behavior is modelled using a Prony series with the dimensionless moduli $g_i$, relaxation times $\tau_i$, and the short-term Neo-Hookean constant $C_{10}^0$, the linearized apparent modulus of beams 1 and 2 are given by[20]:

$$E_1 = 6C_{10,1}^0 \left(1 - \sum_{i=1}^{n} g_{i,1}\left(1 - e^{-t/\tau_{i,1}}\right)\right) \tag{9}$$

$$E_2 = 6C_{10,2}^0 \left(1 - \sum_{i=1}^{n} g_{i,2}\left(1 - e^{-t/\tau_{i,2}}\right)\right) \tag{10}$$

where the additional subscripts 1 and 2 refer to beams 1 and 2, respectively. When one single term of the Prony series is used, the parameter $A$ can be calculated as:

$$A = \frac{E_1}{E_2} = \frac{C_{10,1}^0 \left(1 - g_{1,1}\left(1 - e^{-t/\tau_{1,1}}\right)\right)}{C_{10,2}^0 \left(1 - g_{1,2}\left(1 - e^{-t/\tau_{1,2}}\right)\right)} \tag{11}$$

For the two extreme cases of extremely high strain rates (*i.e.*, $\dot{\varepsilon} \to \infty$) and exteremly low strain rates (*i.e.*, $\dot{\varepsilon} \to 0$), the parameter $A$ is given as:

$$A^{\dot{\varepsilon} \to \infty} = \frac{C_{10,1}^0}{C_{10,2}^0} \tag{12}$$

$$A^{\dot{\varepsilon} \to 0} = \frac{C_{10,1}^0(1 - g_{1,1})}{C_{10,2}^0(1 - g_{1,2})} \tag{13}$$

If we define $\alpha = \frac{C_{10,1}^\infty}{C_{10,2}^\infty}$ and $\beta = \frac{1-g_{1,2}}{1-g_{1,1}}$, the ratio of the elastic moduli corresponding to the extremely high and extremely low strain rates can be re-written as $A^{\dot{\varepsilon} \to 0} = \alpha$ and $A^{\dot{\varepsilon} \to \infty} = \alpha\beta$, respectively. The parameters $\alpha$ and $\beta$ represent the mechanical properties of both beams. Equations (12), (13), and (8) can then be used to determine the buckling direction of a bi-beam made from two visco-hyperelastic beams.

### 4.2. Computational models

Computational models of the bi-beams were built in which the constitutive behaviors of both beams were described using visco-hyperelastic material models where the hyperelastic behavior was modeled using the Neo-Hookean model while a single-term Prony series captured the viscous behavior of the beams. The simulations were performed using the viscoelastic solver of Abaqus (Simulia, Providence, United States). The long-term Neo-Hookean parameter for the stiff material was set to $C_{10,2}^\infty = 1$ MPa as a reference and the dimensionless coefficients of the Prony series for the softer material were set to $g_{1,1} = 0.7$ according to the experimental results obtained through compression mechanical testing of IP-Dip bulk specimens. Given that the focus of the current work is on comparing the adiabatic response of the bi-beams with their isothermal response, the relaxation time for both materials were set to the equal values of $\tau_{1,1} = \tau_{1,2} = 1\ s$. The design parameters of the bi-beams were as follows: $2L = 80$ μm, $a = 12.5$ μm and $R = 1, 2, 3$ mm, which are the length, width, and the radius of curvature of the bi-beams. The geometries were discretized such that there were 6 elements trough the width of the clamped sides of the bi-beams. The elements used in our computational models were all four-node bilinear solid elements with a hybrid formulation and a constant pressure (CPE4H in Abaqus). Clamped boundary conditions were applied to both the top and bottom sides of the bi-beams. We then performed a parametric study to evaluate the effects of the material parameters ($C_{10,1}^\infty$ and $g_{1,2}$) on the switching behavior of the bi-beams and studied the effects of the purposefully introduced imperfections (*i.e.*, the radius of the beam curvature) on the boundaries of the envelopes defining different buckling sides. Two representative deformation rates (*i.e.*, very slow and very fast) of 0.08 and 8000 mm/sec were used for all the simulations.

**4.3. 3D printing and experiments**
Bi-beams with weak differential strain rate dependency were fabricated at macro- and microscales using two 3D printing techniques, namely the jetting of multiple UV-curable polymers (Polyjet) and direct laser writing through two-photon polymerization (2PP). Moreover, we designed and fabricated strain rate-dependent mechanisms at both macro and microscales to showcase the potential of the proposed 3D printable strain rate-dependent meta-devices. The same printing technologies as described above were used for the mechanisms.

*Macroscale bi-beams*
We used a Polyjet 3D printer (J750™ Digital Anatomy™, Stratasys, USA) with a layer thickness of 14 μm and an isotropic in-plane voxel size of 40 μm × 40 μm for the jetting of two commercially available photopolymerizable soft polymers (*i.e.*, TangoPlus and Agilus, Stratasys, USA) with small differences in their strain rate dependencies. For the small strain rates, TangoPlus and Agilus exhibit similar values of the apparent elastic moduli, while at high strain rates, the apparent value of the elastic modulus of Agilus is higher than that of TangoPlus. The bi-beams were designed by laterally attaching TangoPlus beams to Agilus beams, with an aspect ratio of 6 (height = 60 mm, width = 10 mm, and out of plane thickness = 15 mm). Both ends of the bi-beams were symmetrically cut to ensure low sensitivity to unpurposeful (*i.e.*, pre-existing) geometric imperfections (Figures 2A,B)[20]. Moreover, purposeful geometric imperfections were introduced to the TangoPlus side (*i.e.*, left side) as arcs with different radii of curvature (*i.e.*, 600, 800, and 1000 mm). A custom-made compression frame was 3D printed using a fused deposition modeling (FDM) printer (Ultimaker 2, The Netherlands) and polylactic acid (PLA) filaments to subject the bi-beams to different strain rates.

The bi-beams were compressed using an electromechanical testing machine (ElectroPuls E10000, Instron, Norwood, MA, USA) equipped with a 10 kN load cell. The gripper mechanism was compressed using a Lloyd mechanical testing machine (LR5K, Lloyd Instruments, UK) equipped with a 5 kN load cell. A digital camera (Sony a7R with a Sony E 3.5/30 microlens, Sony, Japan) was used to capture the trends of the deformations taking place during the compression tests.

To evaluate the week strain rate dependency in the bi-beams made using Poly-jet printing and PDMS photo curing, we tested two disks made of TangoPlus and Ailus ($D = 28.5$ mm and $h = 12.5$ mm) and two PDMS cubes with edge sizes equal to 40 μm. The Poly-jet printed specimens were tested at $\dot{\varepsilon} = 0.017$ and $\dot{\varepsilon} = 16.7$ 1/s while PDMS specimens were tested similar to the micro-scale specimens made from IP-Dip (see the following sub-section).

*Microscale bi-beams*
Microscale bi-beams were fabricated with submicron resolution using a 2PP 3D printer (Nanoscribe GT2, Nanoscribe GmbH, Karlsruhe, Germany). The fabrication process was performed on a fused silica substrate, which was cleaned by isopropyl alcohol (IPA) and treated with $O_2$ plasma to improve its adhesion to the photocured polymers. A droplet of a commercially available acrylate-based resin (IP-Dip, Nanoscribe, Germany) was drop-casted onto the substrate. The bi-beams were fabricated layer-by-layer with the adjustments of the laser power in each layer of the bi-beam. This adjustment was done to avoid polymerizing each component of the bi-beam separately, which would result in the shadowing of one beam of the bi-beam construct over the neighboring beam. To improve the adhesion of the bottom side of the bi-beams to the silica substrate, each bi-beam was printed on top of a pedestal section made of a similar material (Figure 2C,D). Moreover, a brace structure was fabricated around the bi-beams to enable the consistent application of a uniaxial compressive force to the bi-beams and to prevent the undesired deflections that might otherwise take place during the development phase that follows the printing process. Bi-beams with an aspect ratio 6.4 and with different heights (160, 80, and 40 μm) were fabricated at a constant scanning speed of 10 mm/s (Figure 3A). The horizontal slicing was 0.2 μm and vertical slicing was 0.3 μm. The difference between the apparent elastic moduli of both beams was achieved through the adjustment of the laser power (*i.e.*, 40% *vs.* 45% of a mean nominal laser power of 50 mW). A purposeful geometric imperfection analogous to what was described above for the macroscale bi-beams was introduced to the microscale bi-beams in the form of a slight curvature (radius of curvature: 1-16 mm). The entire assembly was developed in propylene glycol methyl ether acetate (PGMEA) for 25 minutes, followed by immersion in IPA for 5 minutues, and blow-drying with compressed air.

A similar procedure was used to fabricate microscale bi-beams using UV-curable polydimethylsiloxane resin (PDMS, UV-PDMS KER4690, Shin-Etsu Chemical Co., Tokyo, Japan). In the PDMS micro bi-beams, the arrangement of the stiff and less stiff beams was similar to the macroscale bi-beams and without any brace or pedestal structures. To achieve dissimilar strain rate dependencies in each half of the PDMS bi-beams while avoiding shadowing, the laser power was changed during the writing of each layer. For material 2 (Figure 2A,B), a laser power of 100% was used while the laser power was set to 80% for material 1 (100% laser power = 50 mW). The scanning speed was kept constant at 200 μm/s, with a slicing and a hatching distance of 0.2 μm and 0.3 μm, respectively.

We used focused ion beam (FIB) milling to inspect the geometry of the micro bi-beams and their internal structure. This was performed using a Helios microscope (FEI, Helios G4 CX dual beam workstation, Hillsborough, USA), which was operated with at an acceleration voltage of 10 kV and a current of 50 pA.

To characterize the mechanical behavior of the bulk materials from which the IP-Dip bi-beams were made, we manufactured a number of photocured bulk specimens with different geometries and resolutions and tested them under compression loads applied with low and high speeds (i.e., $\dot{\varepsilon} < 0.002$ and $\dot{\varepsilon} = 8.33$ 1/s). We used one specimen to evaluate the strain rate-dependent properties of each geometry. These specimens had different cross-sectional shapes with the same area at 3 different scales (Table 1, Figure 4). The same hatching and slicing parameters (*i.e.*, 0.2 μm and 0.3 μm, respectively) were used. The laser power used for manufacturing the bulk specimens were 40% and 45% of a mean nominal laser power of 50 mW corresponding to materials 1 and 2 in the bi-beams made from IP-Dip respectively.

The microscale specimens were mounted on scanning electron microscope (SEM) stubs and underwent compression tests using a micro-mechanical force sensing probe (FT-NMT03, Femtools, Buchs, Switzerland) either mounted inside a SEM (JSM-IT100LA, JEOL, Tokyo, Japan) or under an optical microscope (Femtools AG, Buchs, Switzerland) depending on the bi-beam size. The samples were gold sputtered with ~40 nm of gold using a sputter coater (JFC1300, JEOL, Japan) prior to the compression tests and SEM imaging. The silicon probe used to compress the bi-beams had a square geometry (50 × 50 μm$^2$) with a force sensing range of up to 200 mN. For bi-beams with a height of 160 µm, compression speeds between 0.1 µm/s (slow) and 500 µm/s (fast) were used to probe the strain rate dependency of the bi-beams. The trend of the buckling was tracked using either the SEM or optical microscope. For fast compression speeds, the probe tip was initially placed 10 µm away from the top of the bi-beams and was accelerated towards it so that it could reach the maximum compression speed before contacting the bi-beams. Similarly, for bi-beams with heights of 80 µm and 40 µm, the fast compression speeds were respectively 250 µm/s and 125 µm/s so as to maintain the same strain rate. For bulk pillars, the fast compression speeds were 500 µm/s (60 µm high pillars), 250 µm/s (40 µm high pillars), and 125 µm/s (20 µm high pillars). The slow compression speed was 0.1 µm/s. All compression tests were limited to ≈ 6-10% of strain to avoid breakages. The force-displacement data was collected for the micro bi-beams as well as for the bulk specimens (Table 1).

*Strain rate-dependent mechanisms*
The mechanisms were designed to open or close depending on the applied strain rate and employed a bi-beam as their strain rate-dependent design element. The force-displacement curves associated with different strain rates were also measured. In the case of the microscale specimens, the measurement protocol was similar to the one described above for the mechanical testing of bi-beams and included the use of the same micro-mechanical force sensing probe. As for the macroscale mechanisms, we used a Lloyd mechanical testing machine (LR5K).

**Table 1.** The geometries and dimensions of the bulk specimens.

| Geometry | small (height = 20 μm) | medium (height = 40 μm) | large (height = 60 μm) |
|---|---|---|---|
| Circle | diameter = 13.5 μm | diameter = 27 μm | diameter = 40.5 μm |
| Rectangle | 8 × 18 μm² | 16 × 36 μm² | 25 × 54 μm² |
| Square | 12 × 12 μm² | 24 × 24 μm² | 36 × 36 μm² |

FIGURE CAPTIONS

**Figure 1.** Mechanics of the bi-beams made by the lateral attachment of two beams with a differential strain rate-dependent response. (A) Bi-beam design where a geometric artifact in the form of a curvature is introduced to favor buckling towards a specific direction (B-C) Computational analysis shows that, in contrast to geometrically-perfect bi-beams that require a 'strong' differential response for switching their buckling direction, strain-rate dependent switching of the buckling direction can be realized with a 'weak' differential response as long as a rationally designed geometric imperfection (in the form of a curvature) is introduced to the compliant side of the bi-beams. (D) A pivot-pivot bi-beam modeled using Euler-Bernoulli beam theory was used to analytically describe the switching behavior of the visco-hyperelastic bi-beams. (E) The master curve of the bi-beams where all the possible combinations of the geometric and material parameters that lead to strain rate-dependent switching of the buckling direction. (F) The prediction of switchability in the buckling direction of bi-beams using nonlinear computational analysis and analytical models. There are two regions (blue and yellow triangles) for which the desired switching behavior can be realized.

**Figure 2.** Additive manufacturing of strain rate-dependent bi-beams. (A) Polyjet multimaterial printing and 2PP were used to respectively fabricate macro- and microscale bi-beams. The stress strain curves show the weak differential strain rate-dependent properties of PDMS cured using 80% and 100% of the full laser power. (B) The switching behavior of the compliant bi-beams ($R = 800$ mm, $2L = 10$ mm, $a = 12.5$ mm) made of TangoPlus-Agilus (left-right) using polyjet printing and PDMS microscale bi-beams ($R = 1$ mm, $2L = 160$ μm, $a = 25$ μm) made using 2PP. (C) The stress-strain curves of IP-Dip bulk materials cured at 40% and 45% of the full laser power using 2PP. The lower laser power results in higher viscoelastic properties. A brace structure was fabricated simultaneously to ensure the uniaxial compression of IP-Dip bi-beams. (D) The switching behavior of IP-Dip bi-beams ($R = 12$ mm, $2L = 160$ μm, $a = 25$ μm) achieved by increasing the compressive strain rate.

**Figure 3.** Effects of the radius of curvature on the reliability of the switching behavior. (A) The SEM images of full-size PDMS microscale bi-beams exhibiting the expected switching behavior. The SEM images clearly show the challenges of producing the required details in PDMS constructs. (B) The effects of scaling on the switching behavior of IP-Dip microscale bi-beams. The SEM images of three different scales of bi-beams indicate the negative effect of scaling on the switching behavior of quarter-size microscale bi-beams. For larger sizes (*i.e.*, full- and half-size microscale bi-beams), the radius of curvature can fine-tune the switching behaviors to ensure reliable buckling behaviors.

**Figure 4.** The effect of the printing power and resolution on the mechanical properties of IP-Dip. (A) The long term and instantaneous Young's moduli of the bulk specimens made from IP-Dip with different geometries and sizes, and cured using 40% and 45% of the full laser power. (B) The effects of the hatching distance and laser power on the programmed switching behavior in quarter-size bi-beams (IP-Dip). The slicing size was set to 300 nm. (B) The SEM images highlight the effects of the bi-beam dimensions on the manufacturability of the geometrical details of the quarter-size microscale bi-beams. The cut-outs present the clamping site of the bi-beams were not reproduced in the quarter-size specimens. (D) The effects of the hatching and slicing sizes on programming the switching behavior of quarter-size microscale bi-beams confirm the weak predictability of the buckling behavior of quarter size bi-beams. (E) SEM images clearly show the damages at the bottom clamp site of bi-beams compressed at 500 μm/s and 0.1 μm/s.

**Figure 5.** Strain rate-dependent compliant grippers. (A) Bi-beams made using compliant and stiff polymers can be used as the strain rate responsive component that determines the opening and grasping functions of compliant grippers. (B) The high-speed and low-speed actuation of macroscale and microscale grippers. (C) The force-displacement curves obtained using the fast and slow compression tests of the macro- and microscale grippers.

**Figure 1**

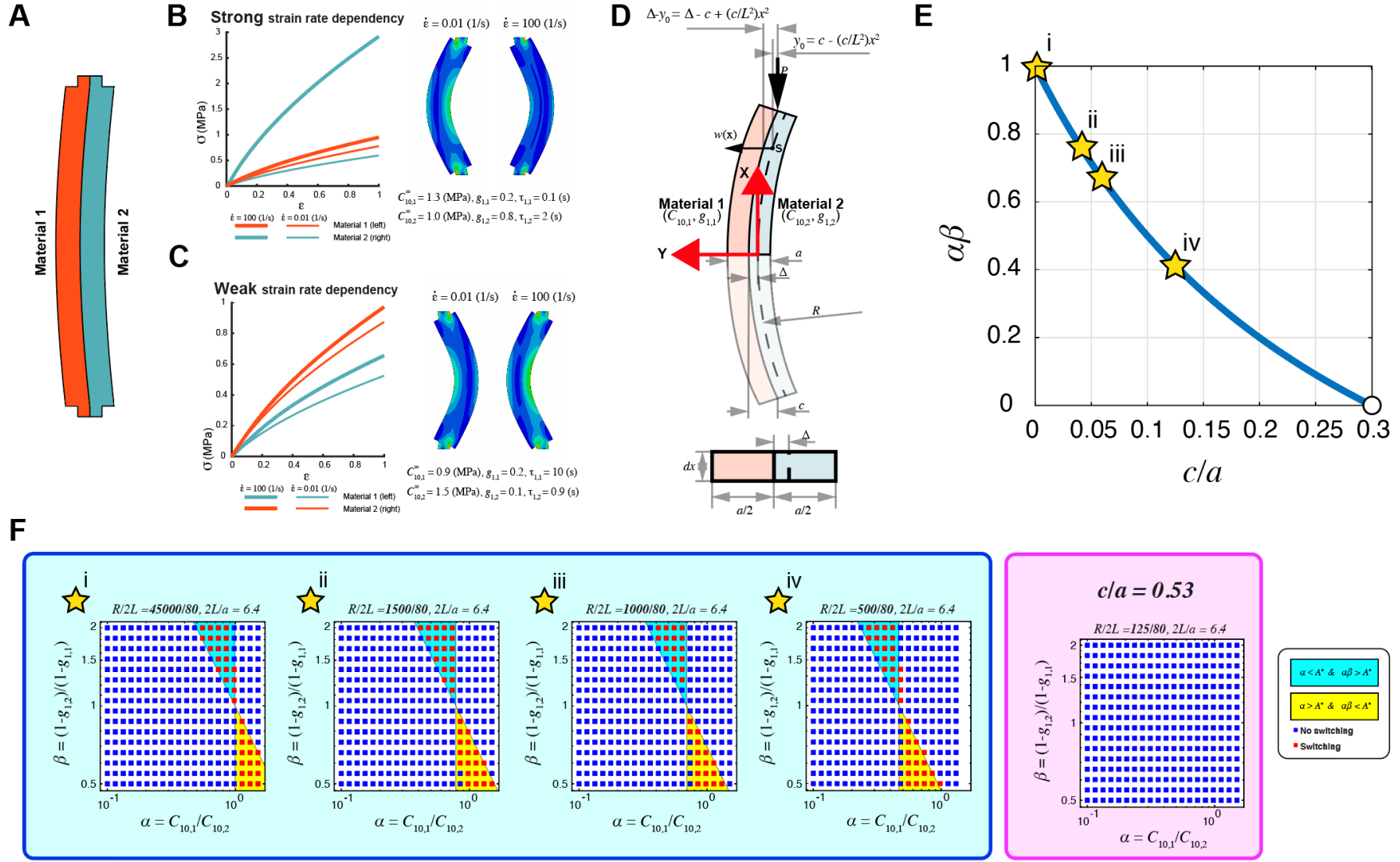

**Figure 2**

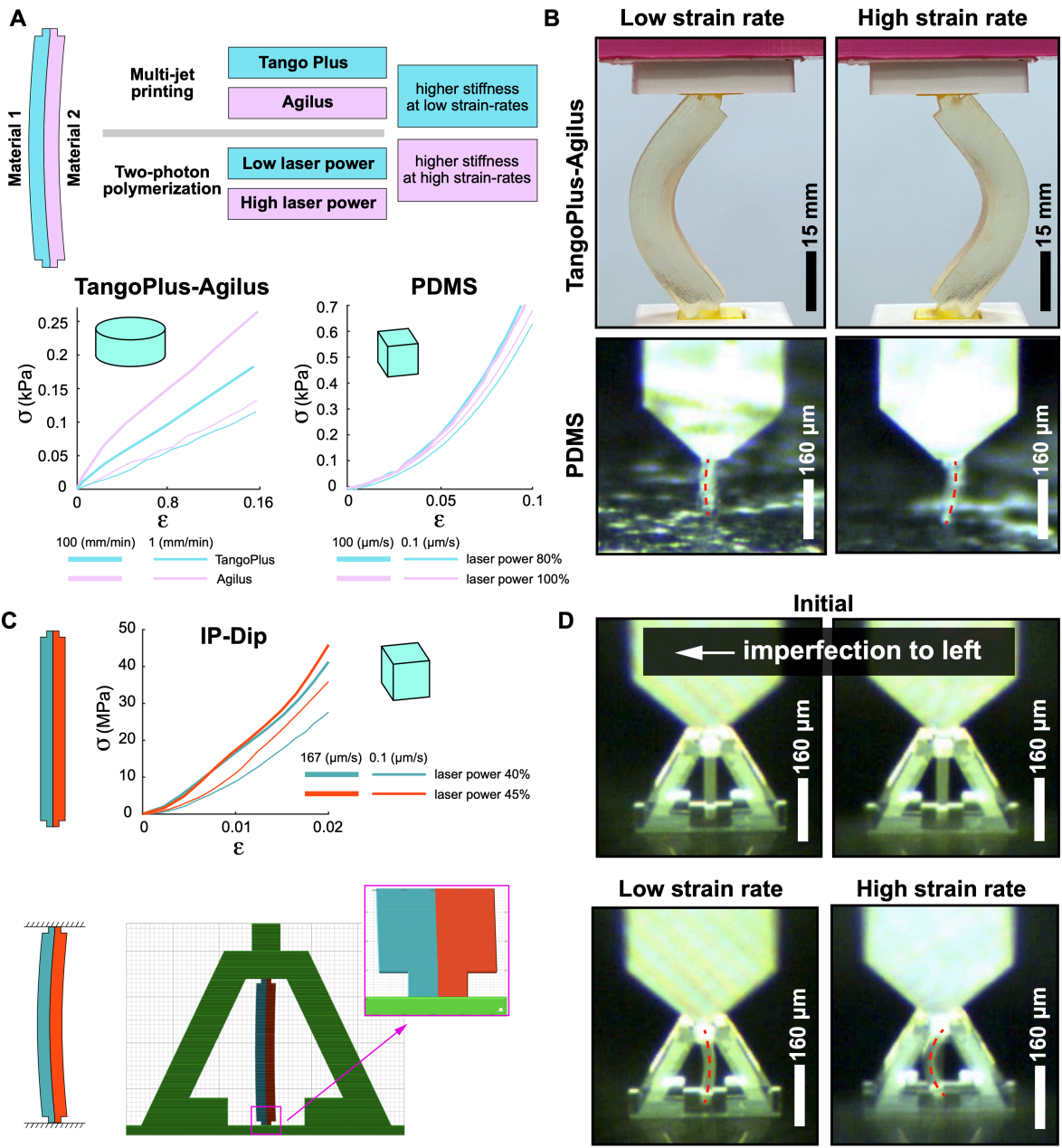

# Figure 3

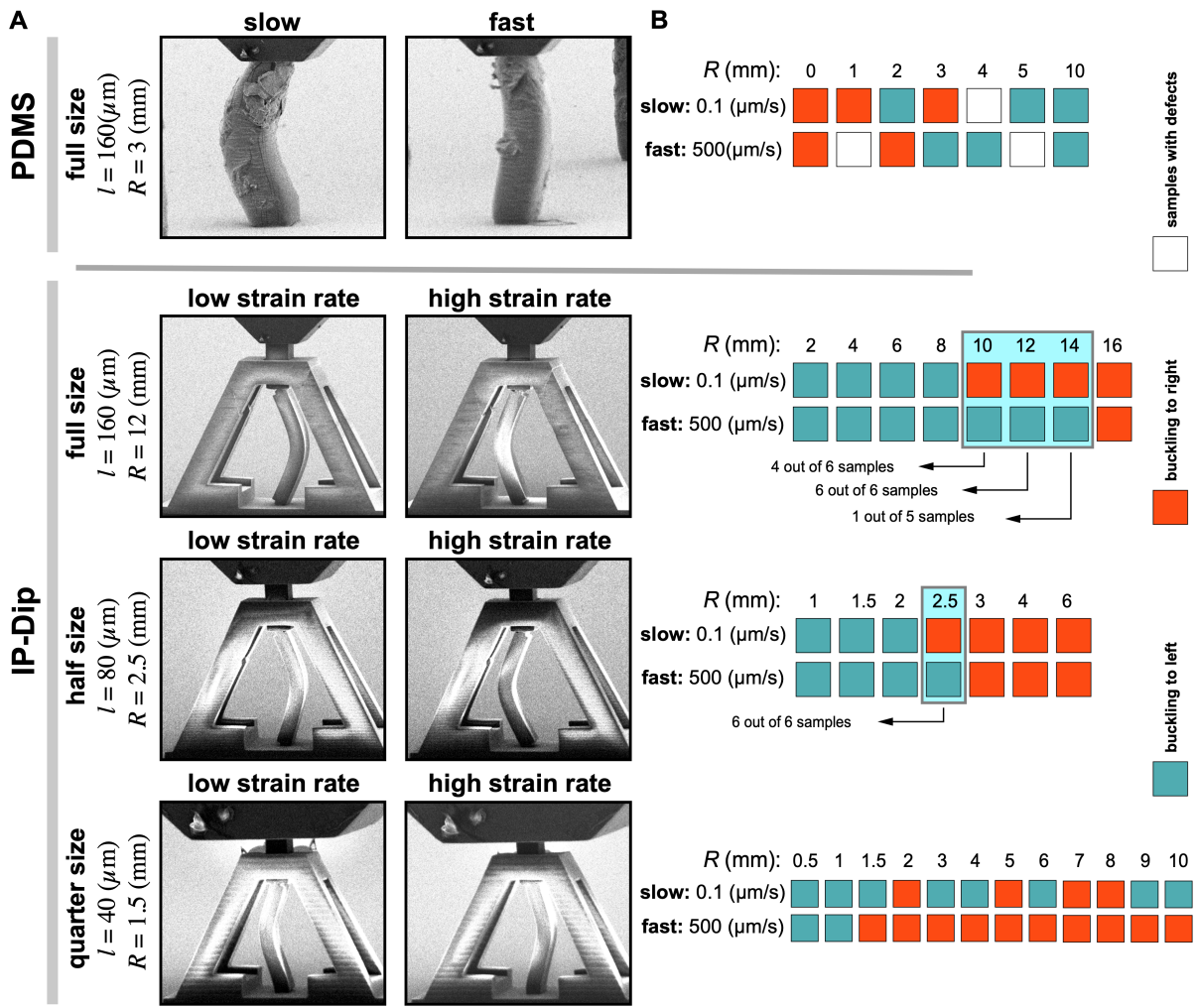

**Figure 4**

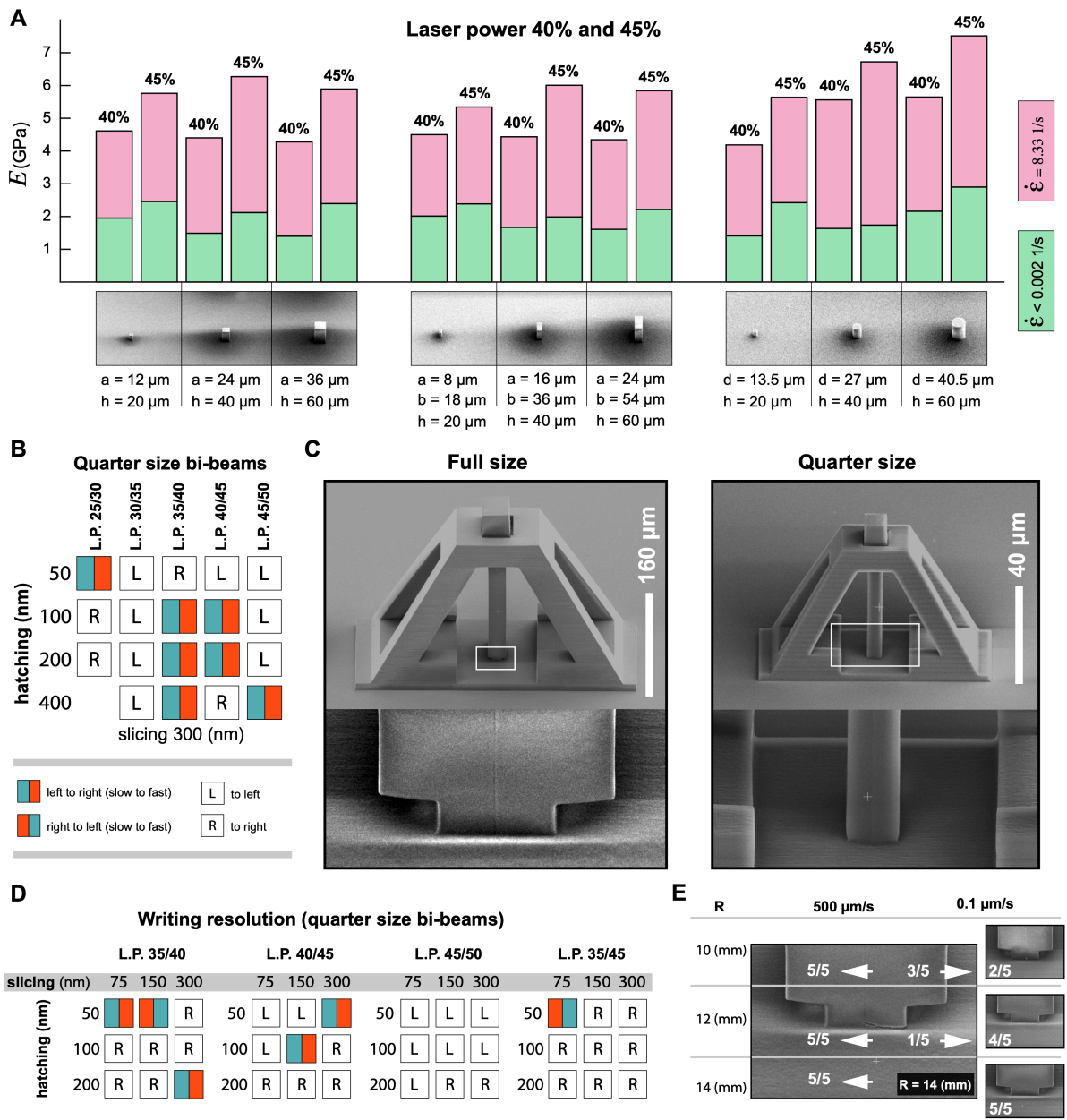

**Figure 5**

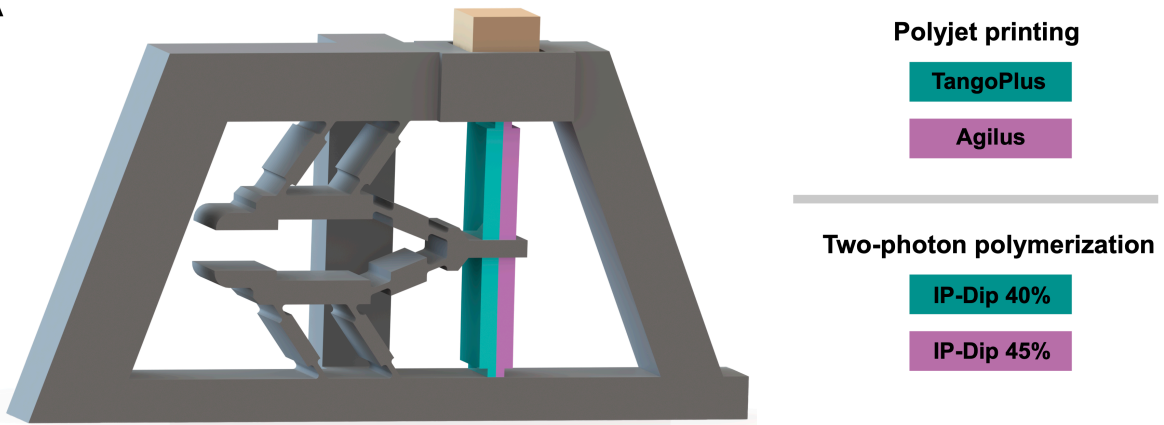

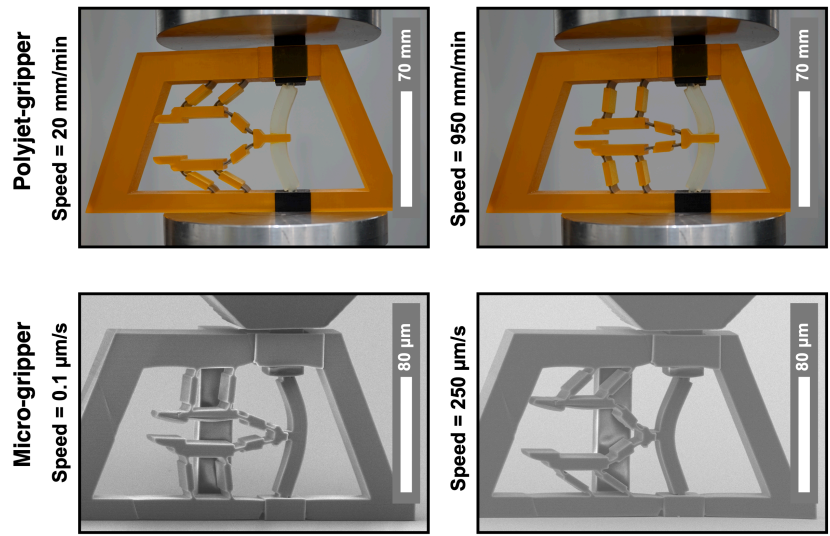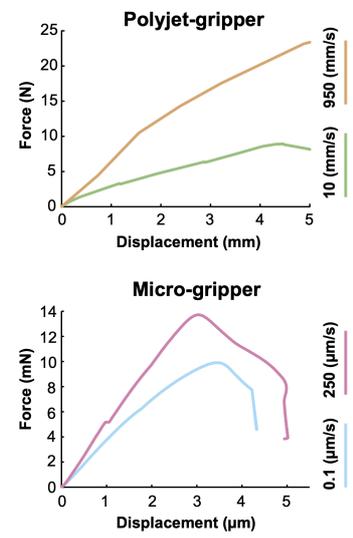